\documentclass[aps,prl,twocolumn,showpacs,superscriptaddress,preprintnumbers,amsmath,amssymb]{revtex4}
\usepackage{graphicx}
\usepackage{dcolumn}
\usepackage{bm}
\usepackage{color}

\begin{document}

\title{Recovery of high-energy photoelectron circular dichroism through Fano interference}

\author{G.~Hartmann}
\affiliation{Institut f\"ur Physik und CINSaT, Universit\"at Kassel, Heinrich-Plett-Str. 40, 34132 Kassel, Germany}

\author{M.~Ilchen}
\affiliation{Institut f\"ur Physik und CINSaT, Universit\"at Kassel, Heinrich-Plett-Str. 40, 34132 Kassel, Germany}
\affiliation{European XFEL GmbH, Holzkoppel 4, 22869 Schenefeld, Germany}

\author{Ph.~Schmidt}
\affiliation{Institut f\"ur Physik und CINSaT, Universit\"at Kassel, Heinrich-Plett-Str. 40, 34132 Kassel, Germany}

\author{C.~K\"{u}stner-Wetekam}
\affiliation{Institut f\"ur Physik und CINSaT, Universit\"at Kassel, Heinrich-Plett-Str. 40, 34132 Kassel, Germany}

\author{C.~Ozga}
\affiliation{Institut f\"ur Physik und CINSaT, Universit\"at Kassel, Heinrich-Plett-Str. 40, 34132 Kassel, Germany}

\author{F.~Scholz}
\affiliation{Deutsches Elektronen-Synchrotron (DESY), Notkestrasse 85, 22607 Hamburg, Germany}

\author{J.~Buck}
\affiliation{Deutsches Elektronen-Synchrotron (DESY), Notkestrasse 85, 22607 Hamburg, Germany}
\affiliation{Institut f\"{u}r Experimentelle und Angewandte Physik, Universit\"{a}t Kiel, Leibnizstr. 19, 24118 Kiel, Germany}

\author{F.~Trinter}
\affiliation{Deutsches Elektronen-Synchrotron (DESY), Notkestrasse 85, 22607 Hamburg, Germany}
\affiliation{Molecular Physics, Fritz-Haber-Institut der Max-Planck-Gesellschaft, Faradayweg 4, 14195 Berlin, Germany}

\author{J.~Viefhaus}
\affiliation{Deutsches Elektronen-Synchrotron (DESY), Notkestrasse 85, 22607 Hamburg, Germany}
\affiliation{Helmholtz-Zentrum Berlin (HZB), Albert-Einstein-Str. 15, 12489 Berlin, Germany}

\author{A.~Ehresmann}
\affiliation{Institut f\"ur Physik und CINSaT, Universit\"at Kassel, Heinrich-Plett-Str. 40, 34132 Kassel, Germany}

\author{M.~S.~Sch\"{o}ffler}
\affiliation{Institut f\"{u}r Kernphysik, J.W. Goethe-Universit\"{a}t, Max-von-Laue-Strasse 1, 60438 Frankfurt am Main, Germany}

\author{A.~Knie}
\affiliation{Institut f\"ur Physik und CINSaT, Universit\"at Kassel, Heinrich-Plett-Str. 40, 34132 Kassel, Germany}

\author{Ph.~V.~Demekhin}\email{demekhin@physik.uni-kassel.de}
\affiliation{Institut f\"ur Physik und CINSaT, Universit\"at Kassel, Heinrich-Plett-Str. 40, 34132 Kassel, Germany}

\begin{abstract}
It is commonly accepted that the magnitude of a photoelectron circular dichroism (PECD) is governed by the ability of an outgoing photoelectron wave packet to probe the chiral asymmetry of a molecule. To be able to accumulate this characteristic asymmetry while escaping the chiral ion, photoelectrons need to have relatively small kinetic energies of up to a few tens of electron volts. Here, we demonstrate a substantial PECD for very fast photoelectrons above 500~eV kinetic energy released from methyloxirane by a participator resonant Auger decay of its lowermost O $1s$-excitation. This effect emerges as a result of the Fano interference between the direct and resonant photoionization pathways, notwithstanding that their individual effects are negligibly small. The resulting dichroic parameter has an anomalous dispersion, i.e. it changes its sign across the resonance, which can be considered as an analogue of the Cotton effect in the X-ray regime.
\end{abstract}


\maketitle

Photoelectron circular dichroism (PECD) is a fundamental chiroptical effect causing a forward-backward asymmetry in the laboratory-frame angular distribution of photoelectrons emitted from chiral molecules in the gas phase. PECD was first predicted theoretically for one-photon ionization \cite{Ritchie1,Ritchie2,Ritchie3} and then verified in pioneering experiments with circularly polarized synchrotron radiation \cite{Expt1,Expt2}. The effect persists also in the multiphoton ionization regime using intense laser pulses \cite{Lux12,Lehmann13}. Because PECD is a pure  electric-dipole effect, it is much stronger than the traditional CD in the photoabsorption spectra of chiral molecules \cite{CDbook}. This fact, together with its enantioselectivity, has established PECD as a powerful tool for chiral recognition in the gas phase \cite{CDbook,REV1,REV2,REV3}.

Ritchie \cite{Ritchie1,Ritchie2,Ritchie3} has proven analytically that PECD arises due to an incomplete compensation of the contributions from emitted partial electron waves with different projections  $\pm m$ of the carried angular momentum $\ell$. Such an inequivalence occurs only for chiral molecules and has a twofold origin: It is induced by the chiral asymmetries of both, the initial bound and the final continuum  electronic  states  entering the dipole-transition amplitudes. For slow photoelectrons with kinetic energies of a few electron volts, initial and final state contributions to PECD can be comparable \cite{Stener04}. As the electron kinetic energy grows, the photoelectron wave packet escapes the molecular ion quickly and does not have enough time to adopt its chiral asymmetry. In other words, both, the chiral initial state and the chiral potential of the ion can be considered almost symmetric for very fast photoelectrons. As a consequence, at high kinetic energies, typically at a few tens of electron volts \cite{Stener04,Powis00} and in extreme cases at about  70~eV \cite{Hergenhahn04}, PECD vanishes.

In the present work, we demonstrate a very general mechanism  of PECD recovery  for electrons with  kinetic energies of as high as a few hundreds of electron volts. This scenario involves an intermediate electronic resonance, which enables an excited metastable wave packet to accumulate chiral asymmetry during its natural decay lifetime, and it does not imply an explicit upper bound for the kinetic energy of a photoelectron. In the vicinity of a resonant excitation, the total transition amplitude for the population of a selected ionic state via the emission of the partial photoelectron continuum wave $\varepsilon\ell m$ is given by the well-known Kramers-Heisenberg formula (atomic units are used throughout):
\vspace{-0.1cm}
\begin{equation}
\mathcal{D}_{\ell m k}({ \delta})=d_{\varepsilon\ell m k} +\frac{V_{\varepsilon\ell m}D_k}{{ \delta} +i\gamma}. \label{eq:FANO}
\end{equation}
Here, ${ \delta}=\omega-E_r$ is the detuning of the exciting-photon energy $\omega$ from the resonance energy $E_r$ and  $\gamma=\Gamma_r/2$ is  half of the respective decay width $\Gamma_r$. The photoelectron kinetic energy $\varepsilon$ is related to the ionization potential ($IP$) of the final ionic state via $\varepsilon=\omega-IP$.

Equation~(\ref{eq:FANO}) describes the Fano interference \cite{Fano61} between the amplitude $d_{\varepsilon\ell m k}$ for the direct ionization of a selected ionic state (here $k=0,\pm1$ stands for the light polarization) and the transition amplitude for the resonant excitation (represented  by the dipole matrix element $D_k$) and subsequent electronic decay into the same final state of the `ion + photoelectron' (represented by the Coulomb matrix element $V_{\varepsilon\ell m}$). The Fano interference is responsible for the shape of a resonant profile in the total photoionization cross section across an intermediate excited state \cite{Fano61}. For a high-energy inner-shell excitation and subsequent resonant Auger decay, the contribution of the direct ionization channel is usually small compared to the resonant pathway. As a consequence, the Fano interference is rather weak, and the respective profile in the angle-averaged cross section has the form of an almost symmetric peak on a small constant background.

The angle-resolved one-photon ionization spectrum of randomly-oriented molecules induced by circularly polarized light is given by the following standard formula for the differential  cross section \cite{Cherepkov81}
\vspace{-0.1cm}
\begin{equation}
\frac{\mathrm{d}\sigma^\pm}{\mathrm{d\Omega}}=\frac{ {\sigma} }{4\pi}\left[ 1 \pm \beta_1 P_1(\cos \theta)  -\frac{1}{2} \beta_2 P_2(\cos \theta)\right]. \label{eq:DPICS}
\end{equation}
Here, $\sigma$ is the total photoionization  cross section, $\beta_1$ is the dichroic parameter which describes PECD in chiral molecules, $\beta_2$ is the anisotropy parameter, `$\pm$' stands for the positive and negative helicity of the circularly polarized radiation, $P_L(\cos \theta )$  are  the Legendre polynomials, and $\theta$ is the angle between the direction of the propagation of the ionizing radiation and the direction of the emission of photoelectrons.

In general, the dichroic $\beta_1$ and anisotropy $\beta_2$ parameters are given by  coherent superpositions  \cite{Cherepkov81,Ilchen17} of the transition amplitudes $\mathcal{D}_{\ell m k}({ \delta})=\mathcal{D}_{j}$
\begin{equation}
 \begin{array}{c}\beta_L =\sum_{pj} b^L_{pj} \mathcal{D}^\ast_p \mathcal{D}_j, \end{array} \label{eq:Bettas}
\end{equation}
where $b^L_{pj}$ are known kinematic coefficients. The impact of Fano interferences on anisotropy parameters $\beta_2$ of high-energy photoelectrons in molecules is well documented \cite{Demekhin09,Demekhin10a,Demekhin10b,Demekhin10c,Knie14,Antonsson15,Knie16,Nandi17}. In the off-resonance excitation regime, anisotropy in the photoemission is governed by the direct ionization channel, while in the on-resonance regime, it is determined by the resonant channel (the first and second terms in the amplitude~(\ref{eq:FANO}), respectively). This induces a rather broad dispersion of the $\beta_2$ parameter across the resonance \cite{Demekhin09,Demekhin10a,Demekhin10b,Demekhin10c,Knie14,Antonsson15,Knie16,Nandi17}. The Fano interference, i.e. the impact of the cross terms of the direct and resonant contributions, causes moderate changes of the dispersion created by incoherent effects of those two contributions \cite{Nandi17}.

The situation changes completely if one considers the dichroic parameter $\beta_1$. In the high kinetic energy regime, the chiral asymmetry parameter provided by the direct photoionization channel is negligibly small: $\beta_1^{Dir}\approx 0$ (see discussion above). In the electric-dipole approximation, the individual dichroic parameter provided by the resonant photoionization channel is strictly zero, $\beta_1^{Res}= 0$, as in the case of the traditional CD in the photoabsorption spectra of chiral molecules. This is because the angular momentum  indices $\ell m$ in the decay transition matrix element in the second term of the amplitude~(\ref{eq:FANO}) are decoupled from the polarization index $k$ in the excitation matrix element, and the respective summations in Eq.~(\ref{eq:Bettas}) can be performed independently. This validates the so-called two-step model  \cite{2ST}, in which an electronic decay of an excited state can be described independently of the initial excitation step.

The contribution of the Fano interference between the direct and resonant photoionization pathways to the dichroic parameter is given by:
\vspace{-0.2cm}
\begin{equation}
\begin{array}{c} \beta_1^{Fano}{  (\delta)} =   \sum_{pj}\end{array}   b^1_{pj} \left\{ d^\ast_p \frac{(VD)_j}{{ \delta} +i\gamma} + d_j \frac{(VD)^\ast_p}{{ \delta} -i\gamma}\right\} . \label{eq:Dichroic}
\end{equation}
In this equation, the angular momentum indices of the photoelectron are still coupled to the polarization of the exciting photon through the direct transition amplitude in each term. Moreover, equation~(\ref{eq:Dichroic}) can straightforwardly be simplified to the following analytic form:
\vspace{-0.2cm}
\begin{equation}
 \beta_1^{Fano} ({ \delta}) = \frac{ \kappa { \delta}+{ \mu} }{{ \delta}^2 +\gamma^2} , \label{eq:AnomDispers}
\end{equation}
with the real coefficients $\kappa$ and ${ \mu}$. Equation~(\ref{eq:AnomDispers}) describes an anomalous dispersion of $\beta_1^{Fano}$ across the resonance: This dichroic parameter changes its sign in a very close proximity of the resonance energy $E_r$  at ${ \delta}=-{ \mu}/\kappa$. This situation is very similar to the anomalous optical rotatory dispersion, also known in the literature as Cotton effect \cite{Cotton1,Cotton2,Cotton3}. {The Fano interference between the overlapping bright and dark plasmonic modes is known to be responsible for an enhancement of the traditional CD in the photoabsorption of chiral nanostructures in the visible and infrared spectral ranges \cite{Nano1,Nano2}.} A sizable influence of electron correlation processes on the dichroic parameter in the vicinity of an autoionizing resonance has also been observed for chiral metal-organic complexes at moderate photoelectron kinetic energies of about 55~eV \cite{Catone12}. The present theoretical analysis suggests a { novel} mechanism of recovery of PECD in the vicinity of high-energy resonant excitation through the Fano interference, where this chiroptical effect due to the direct ionization of valence electrons is expected to be negligibly small.

In order to make an estimate of the magnitude of the proposed effect, we considered  the participator resonant Auger decay  of the lowermost O $1s$-excitation in the prototypical methyloxirane molecule, for which one-photon PECD has been studied extensively \cite{Stener04,MOX1,MOX2,MOX3,MOX4,MOX5,MOX6}. The suggested resonant excitation appears in the ionization yield as a broad peak with the maximum at the photon energy of  $\omega=535.11$~eV \cite{MOX3}. We consider the participator resonant Auger decay in the ground and first excited electronic states of the molecular ion. The photoelectron signals associated with the ionization of HOMO and HOMO-1 of methyloxirane emerge in the  spectrum  as slightly overlapping vibronic bands centered at the binding energies of about 10.4 and 11.4~eV, respectively, and they are separated well from the rest of the spectrum \cite{MOX2}. We, thus, investigate PECD for electrons with kinetic energies of around $\varepsilon \approx 524$~eV.

The participator resonant Auger decay of methyloxirane was described by an {\it ab initio} theoretical approach developed in our previous angle-resolved studies of core-excited/ionized diatomic \cite{Demekhin09,Demekhin10a,Demekhin10b,Demekhin10c}, polyatomic \cite{Knie14,Antonsson15,Knie16,Nandi17}, and chiral \cite{Ilchen17,MOX6} molecules. The electronic transition amplitudes were computed by the stationary Single Center (SC) method \cite{SC0,SC1,SC2}, which enables an accurate description of partial electron continuum waves in
molecules. Calculations were performed in the relaxed-core Hartee-Fock approximation at the equilibrium internuclear geometry of the electronic ground state of R(+) methyloxirane \cite{GEOMETRY}, as described in details in Refs.~\cite{Demekhin09,Demekhin10a,Demekhin10b,Demekhin10c,Knie14,Antonsson15,Knie16,Nandi17}.  The SC expansions of all occupied/excited molecular orbitals and of the photoelectron continuum waves were restricted by harmonics with $\ell,\vert m\vert \leq 50$ and $35$, respectively. The angle-resolved participator Auger decay spectra were computed for the whole vibrational band of each final ionic state. In order to obtain such vibrationally unresolved decay spectra, we applied an analytical procedure described in the appendix of Ref.~\cite{Knie14}. It requires only the shape of the excitation spectrum of the resonance and keeps the Fano interference unchanged. The respective excitation spectrum was approximated by the Gaussian function with a $\mathrm{FWHM}=2$~eV \cite{MOX3}.

\begin{figure}
\includegraphics[scale=0.89]{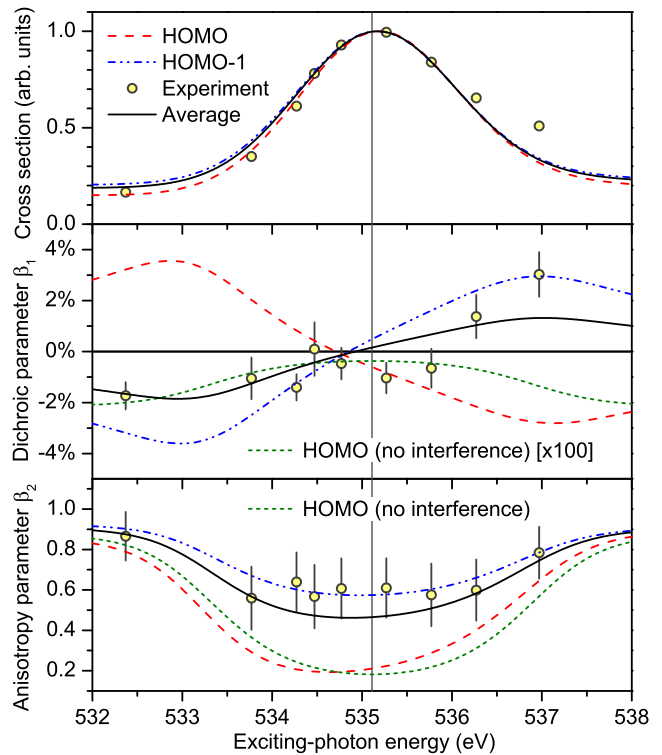}
\caption{Relative cross sections (upper panel) for the population of the HOMO and HOMO-1 ionized  final states (see legend) together with the respective dichroic (middle panel) and anisotropy (lower panel) parameters computed in the present work as functions of the exciting-photon energy in the vicinity of the lowermost O $1s$-excitation of R(+) methyloxirane. The energy position of the resonance $E_r=535.11$~eV \cite{MOX3} is indicated by the vertical solid line. {Dichroic parameters are given in percent of the respective total cross sections, which} are normalized to unity at their maxima. The dichroic and anisotropy parameters, computed for the HOMO  ionized final state without including the Fano interference, are depicted by the green dotted curves. Note that this dichroic parameter is shown on an enhanced scale, as indicated by the factor $[\times 100]$ in the legend. Open circles with error-bars represent the present experimental data for S($-$) methyloxirane (dichroic parameter is shown with the opposite sign) measured for the two unresolved HOMO and HOMO-1 electron bands (see also text). Solid curves represent average theoretical results obtained with a 30\% of HOMO to 70\% of HOMO-1 ratio.}\label{fig:Theory}
\end{figure}

Results of the present calculations are summarized in Fig.~\ref{fig:Theory}. As one can see from the upper panel of this figure, the direct ionizations of HOMO and HOMO-1 electrons at the considered exciting-photon energies are rather substantial (note a constant background in each photoionization cross section). Nevertheless, the influence of the Fano interference on $\sigma(\omega)$ is moderate: The resulting profiles have a form of slightly asymmetric peaks. For the HOMO ionized final state, for instance, the presently computed on- and off-resonance anisotropy parameters are equal to: $\beta_2^{Dir} = 0.90$ and $\beta_2^{Res} = 0.04$. Their incoherent incorporation (no interference) induces a very broad dispersion of the computed anisotropy parameter across the considered core excitation (green dotted curve in the  lowermost panel of  Fig.~\ref{fig:Theory}). The Fano interference itself slightly changes the computed  $\beta_2(\omega)$ dispersion (cf., red dashed and green dotted curves in the lowermost panel of the figure). Similar conclusions apply for the computed anisotropy parameter of the HOMO-1 ionized final state (blue dash-dotted curve in the panel).

We now turn to the discussion of the respective dichroic parameters depicted in the  middle panel of Fig.~\ref{fig:Theory}. The off-resonance dichroic parameter, computed in the present work for the  HOMO ionized state, is equal to $\beta_1^{Dir} = -0.022\%$, and it is relatively small due to the very high kinetic energy of the photoelectron. As justified above, in the electric-dipole approximation implied here, the on-resonance dichroic parameter vanishes. Without Fano interference, this dichroic parameter would have a regular dispersion which follows individual trends in the on- and off-resonance excitation regimes (green dotted curve in the middle panel; note also the enhancement factor of 100). By comparing the red dashed and green dotted curves in the  middle panel of Fig.~\ref{fig:Theory}, one can recognize that the Fano interference enhances the computed chiral asymmetry by about two orders of magnitude and induces an anomalous dispersion of the $\beta_1(\omega)$ parameter across the resonance. A similar effect can be observed  for the HOMO-1 ionized final state, but the computed anomalous dispersion of $\beta_1(\omega)$ has an opposite sign (blue dash-dotted curve in the middle panel of Fig.~\ref{fig:Theory}).

\begin{figure}
\includegraphics[scale=0.86]{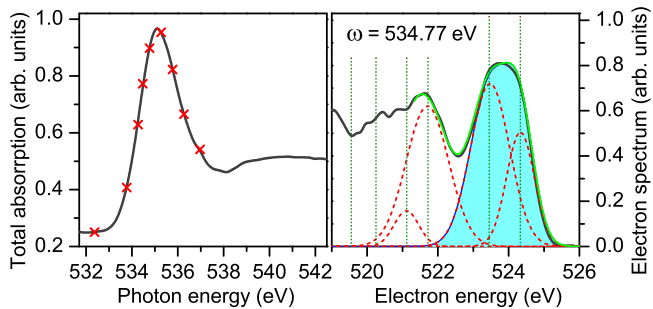}
\caption{Left panel: The presently measured total absorption spectrum of methyloxirane. Exciting-photon energies $\omega$ used to record the angle-resolved photoelectron spectra are marked by crosses. Right panel: A typical  spectrum of electrons recorded at the photon energy of 534.77~eV (dark solid curve) and the present fit of this spectrum around HOMO and HOMO-1 bands (light solid curve).  The energy positions of the HOMO-$n$ electron bands, obtained using binding energies from Ref.~\cite{MOX2}, are indicated by the vertical dotted lines. Breakup of the spectrum into the individual contributions from the HOMO -- HOMO-3 electron bands,  obtained by the present fitting, is also shown by the dashed Gaussian curves. The filled area, used in the present analysis of the electron angular emission distributions, represents unresolved contributions from the HOMO and HOMO-1 electron bands.}\label{fig:Experiment}
\end{figure}

The present theoretical predictions were verified by an experiment at the variable polarization XUV beamline P04 \cite{DESYP04} in the few-bunch mode of the synchrotron radiation facility PETRA~III at DESY in Hamburg, Germany. This beamline uses an APPLE-II-type undulator to cover a large photon energy range from 250~eV to 3~keV with a resolving power higher than $10^4$ and a flux higher than $10^{12}$ photons per second. A polarization analysis of this beamline, performed for the present experimental conditions, led to an estimated degree of circular polarization of $97\pm2\%$. The samples, S($-$) methyloxirane (99\%, Sigma Aldrich) as well as gaseous neon, were introduced into the interaction chamber effusively via a gas needle. At first, an absorption scan was performed in small steps in the vicinity of the lowermost O $1s$-excitation of methyloxirane, and the exciting-photon energy was calibrated to the data of Ref.~\cite{MOX3}. The presently measured  photoabsorption spectrum of methyloxirane is depicted in the left panel of Fig.~\ref{fig:Experiment} by the solid curve.

Thereafter, the angle-resolved electron spectra were recorded for left- and right-handed circularly polarized synchrotron radiation at a few exciting-photon energies across the resonance (marked by crosses on  top of the absorption spectrum in Fig.~\ref{fig:Experiment}). The electrons created in the interaction region were detected with 16 independently-operating time-of-flight spectrometers oriented in a plane rotated by 51.8$^\circ$ with respect to the light propagation direction. This allowed for the determination of electron angular emission distributions with dipole and non-dipole contributions. This setup is described in more detail in Ref.~\cite{CookieBox}. In order to achieve a better spectral energy resolution, a nominal retardation voltage of 510~V was applied to the flight tubes to decelerate fast Auger electrons. The magnetic field in the interaction region was minimized by a set of three pairs of Helmholtz coils.

The detector calibration, i.e. the time-of-flight to kinetic energy conversion and the transmission function as a function of kinetic energy, was performed by measuring a series of neon 2p-photoelectron  spectra for different photon energies. For these spectra, the respective  kinetic energies and anisotropy parameters $\beta_2(\omega)$ are known \cite{TABLES,BECKER}. Importantly, at the employed high photon energies, non-dipole effects cause a noticeable  forward-backward asymmetry in the photoelectron emission \cite{TABLES}. Since this non-dipole asymmetry of the neon 2p-photoelectrons is constant within the narrow energy range across the considered core excitation of methyloxirane \cite{TABLES}, the forward-backward sensitivity of the detectors was calibrated using the known \cite{TABLES} constant value {of the asymmetry} of 3.3\%. In contrast to the presently discussed contribution from PECD, the non-dipole contribution to {the forward-backward asymmetry} of methyloxirane is helicity-independent. As a consequence, this unknown constant contribution cancels out when subtracting two spectra measured for the left- and right-handed circular polarizations (see below).

Unfortunately, we were not able to resolve resonant Auger electrons for the HOMO and HOMO-1 ionization bands of methyloxirane from each other, yet we could resolve those two bands from the rest of the spectrum. For this purpose, the calibrated, transmission- and polarization-corrected electron spectra were fitted with Gaussian curves, which were  positioned at the kinetic energies given through the respective photon energy and the binding energies of Ref.~\cite{MOX2}. The present fitting procedure is illustrated in the right panel of Fig.~\ref{fig:Experiment} on the example of a typical electron spectrum. As one can see, the unresolved contribution from the HOMO and HOMO-1 electron bands (high kinetic energy part of the spectrum highlighted by the filled area)  can unambiguously be separated from the contributions of the other electron bands. Such unresolved contributions, determined at the  chosen photon energies for left- and right-handed circularly polarized light and different emission angles $\theta$,  were used to extract the respective quantities via the parameterizations $\sigma^+(\theta)+\sigma^-(\theta)=2\sigma-\sigma\beta_2P_2(\cos \theta)$  and $\sigma^+(\theta)-\sigma^-(\theta)=2\sigma\beta_1P_1(\cos \theta)$.

Results of the present measurement are depicted in Fig.~\ref{fig:Theory} by open circles. Its upper panel demonstrates a very good agreement between the normalized computed and measured cross sections  $\sigma(\omega)$ (note that the last high-energy point contains a contribution  from the next core-excited resonance, which was not included in the calculations). The lowermost panel of Fig.~\ref{fig:Theory}  demonstrates a very good  agreement between the measured average anisotropy parameter $\beta_2(\omega)$ and that computed for the HOMO-1 band. The experimental uncertainties, indicated by error-bars have three sources: the Poisson statistical counting, fitting errors, and calibration uncertainties. For $\beta_2(\omega)$, detector calibration using the neon 2p-line gave the major contribution (the available data vary from $\beta_2=1.14$ \cite{TABLES} to $\beta_2=1.35\pm 0.11$ \cite{BECKER}). For  $\beta_1(\omega)$, all three sources provide comparable contributions to the experimental uncertainties (note that, since $\beta_2(\omega)$ and $\beta_1(\omega)$ were determined via different parameterizations, calibration uncertainties for the former do not contribute to those for the latter). Similarly to $\beta_2(\omega)$, the measured average dichroic parameter $\beta_1(\omega)$ follows the trend imposed by the parameter computed for the HOMO-1 band (cf. open circles and blue dash-dotted curve in the middle panel of Fig.~\ref{fig:Theory}). To guide the eye, the average theoretical results obtained with 30\% contribution from { the normalized cross section of} HOMO and 70\% from {that of} HOMO-1 bands are shown in Fig.~\ref{fig:Theory} by black solid curves. The applied ratio is in reasonable agreement with previous measurements of Ref.~\cite{MOX3}. Importantly, the present experiment confirms the anomalous dispersion of the dichroic parameter $\beta_1(\omega)$, as predicted theoretically. Even the size of the observed effect is in semi-quantitative agreement with theory.

In conclusion, we demonstrate a dramatic effect of the Fano interference between the direct and resonant ionization pathways on the PECD of high-energy photoelectrons in the vicinity of the lowermost O $1s$-excitation of methyloxirane. The intermediate resonance causes an anomalous dispersion of PECD and enhances the chiral asymmetry by about two orders of magnitude. We justify analytically that this effect of a metastable core-excited state on PECD is general. The present angle-resolved photoionization experiment, performed with circularly polarized synchrotron radiation, unambiguously confirms that the predicted recovery of PECD can be observed even for two unresolved photoelectron bands. { Taking into account the element- and site-selectivity \cite{Ilchen17} of the presently studied effect, PECD after energetically well-separated core excitation and subsequent resonant Auger decay may become a powerful complementary tool for chiral recognition in the X-ray domain. The presented results reveal a significantly broadened applicability of PECD for chiral recognition in the gas phase and suggest universality of this tool with respect to the exciting-photon spectral range and photoelectron kinetic energies.}

\begin{acknowledgements}
This work was funded by the Deutsche Forschungsgemeinschaft (DFG) -- Projektnummer 328961117 -- SFB 1319 ELCH (Extreme light for sensing and driving molecular chirality). M.I. acknowledges funding from the Volkswagen Foundation within a Peter Paul Ewald-Fellowship. We acknowledge DESY (Hamburg, Germany), a member of the Helmholtz Association HGF, for the allocation of synchrotron radiation beamline P04 at PETRA III under proposal I-20170344. We would like to thank Kai Bagschik and J\"{o}rn Seltmann for assistance in using the angle-resolved electron time-of-flight spectrometer. {We thank Rudolf Pietschnig for suggesting to us the analogy with the Cotton effect.} This research was supported in part through the Maxwell computational resources operated at DESY.
\end{acknowledgements}

\end{document}